\documentstyle[twocolumn,pre,aps]{revtex}

\begin{document}

\draft
\input epsf

\title{Propagation of a short laser pulse in a plasma}

\author{Borge Nodland${}^{1,3,4}$ and C. J. McKinstrie${}^{2,3,4}$}

\address{
${}^1$Department of Physics and Astronomy, University of Rochester, 
Rochester, New York, 14627\\
${}^2$Department of Mechanical Engineering, University of Rochester, 
Rochester, New York, 14627\\
${}^3$Rochester Theory Center for Optical Science and Engineering,
University of Rochester, Rochester, New York, 14627\\ 
${}^4$Laboratory for Laser Energetics, University of Rochester, 
Rochester, New York, 14623\\ 
}

\author{
The propagation of an electromagnetic pulse in a plasma is studied for
pulse durations that are comparable to the plasma period. When the
carrier frequency of the incident pulse is much higher than the plasma
frequency, the pulse propagates without distortion at its group speed.
When the carrier frequency is comparable to the plasma frequency, the
pulse is distorted and leaves behind it an electromagnetic wake.
}

\address{PACS numbers: 52.40.Nk, 03.40.Kf, 42.25.Bs}

\date{To be published in Physical Review E {\bf 56} (December 1, 1997)}

\maketitle

\section{Introduction}

The propagation of an electromagnetic wave in a medium \cite{jac} is
controlled by the dielectric function, wich characterizes the response
of the medium to the applied electromagnetic field. The dielectric
function of a plasma is $1 - \omega_p^2/ \omega^2$, where $\omega_p$ is
the plasma frequency, and $\omega$ is the frequency of any Fourier
component of the wave. This simple formula also characterizes the
response of a dielectric medium when the Fourier spectrum of the wave
contains frequencies that are much higher than the resonance
frequencies of the medium.

When a monochromatic wave of frequency $\omega$ is incident upon a
vacuum-plasma boundary, a fraction $2 k_I / (k_T + k_I)$ is
transmitted and a fraction $(k_T - k_I)/(k_T + k_I)$ is reflected,
where $k_I = \omega / c$ is the wave number of the incident wave, and
$k_T = {(\omega^2 - \omega_p^2)}^{1/2} / c$ is the wave number of the
transmitted  wave. Now consider an electromagnetic pulse with carrier
frequency $\omega_c$ and envelope frequency $\omega_e$. The formulas
for the transmission and reflection of a monochromatic wave are also
valid for a long pulse, provided one substitutes $\omega_c$ for
$\omega$. When $\omega_c \leq \omega_p$, the incident pulse is
reflected completely. When $\omega_c > \omega_p$, the transmitted part
of a long pulse propagates without distortion at its group speed $ c
{(1 - \omega_p^2/\omega_c^2)}^{1/2}$. Eventually, the transmitted pulse
disperses. These results are known to be valid for $\omega_e \ll
\omega_p$. In this paper we study electromagnetic propagation in the
complementary regime $\omega_e \sim \omega_p$. Short-pulse
propagation is generally relevant when the long-envelope approximation
is not valid. A specific example is the wakefield accelerator concept
\cite{taj,spr}.  

We use Laplace transform and Green function techniques to analyze the
interaction between the laser pulse and the plasma. We find that the
interaction can be divided into two stages, one in which temporal
transmission and reflection occurs at the vacuum-plasma boundary, and
one in which the transmitted and reflected pulse propagate in the
plasma and vacuum, respectively. We then present details of what
happens at each stage, for incident pulses of varying carrier
frequency and duration.

\section{Analysis}

We consider a laser pulse with electric field $E(t, x)$ that propagates
in vacuum when $x < 0$, enters the plasma at $x = 0$, and propagates
through the plasma for $x > 0$. We assume that the plasma is
characterized by a plasma frequency $\omega_p$. The wave equation
obeyed by $E(t, x)$ is given by

\begin{equation}
(\partial_{tt}^2 - c^2 
\partial_{xx}^2 
+ \omega_p^2) E(t, x) = 0,
\label{eq1}
\end{equation}
where $\partial_{tt}^2$ and $\partial_{xx}^2$ are second-order partial
derivatives with respect to time $t$ and space $x$, and where $c$ is
the speed of light in vacuum.

Let $\omega_p t \rightarrow t$, $\omega_p x/ c \rightarrow x$, so that
$t$ and $x$ become dimensionless. Then the wave equation (\ref{eq1})
becomes

\begin{equation}
(\partial_{tt}^2 -  \partial_{xx}^2 + 1) E(t, x) = 0.
\label{eq1.5}
\end{equation}

In general, some fraction of the incoming laser pulse is reflected at
the vacuum-plasma boundary, while the rest is transmitted into the
plasma. We denote the incident electric field by $E_I(t, x)$, the
reflected field by $E_R(t, x)$, and the transmitted field by $E_T(t,
x)$. Since the electric field is continuous across the boundary \cite{jac},

\begin{equation}
E_I(t, 0) + E_R(t, 0) = E_T(t, 0). 
\label{eq20}
\end{equation}
Similarly, since the magnetic field of the pulse is continuous
across the boundary \cite{jac},

\begin{equation}
\partial_x E_I(t, 0) + \partial_x E_R(t, 0) = \partial_x E_T(t, 0).
\label{eq20.3}
\end{equation}

We next take the temporal Laplace transform of Eqs. (\ref{eq20}) 
and (\ref{eq20.3}) to obtain the equivalent boundary conditions 
in Laplace space,

\begin{eqnarray}
\overline{E}_I(s, 0) + \overline{E}_R(s, 0) 
&=& \overline{E}_T(s, 0), \nonumber \\
\partial_x \overline{E}_I(s, 0) 
+ \partial_x \overline{E}_R(s, 0) 
&=& \partial_x \overline{E}_T(s, 0).
\label{eq21}
\end{eqnarray}

In general, the incident field $E_I(t, x)$ propagates to the right
(toward the plasma), while the reflected field $E_R(t, x)$ propagates
to the left (away from the plasma). We may therefore assume that
$E_I(t, x)$ and $E_R(t, x)$ have the space-time dependencies

\begin{eqnarray}
E_I(t, x) &=& E_I(t - x), \nonumber \\
E_R(t, x) &=& E_R(t + x),  
\label{eq22.3}
\end{eqnarray}
which are consistent with the reduced equations
\begin{eqnarray}
(\partial_t + \partial_x) E_I(t, x) &=& 0, \nonumber \\   
(\partial_t - \partial_x) E_R(t, x) &=& 0.   
\label{eq22.8}
\end{eqnarray}
By taking the temporal Laplace transform of Eqs. (\ref{eq22.8}), 
and letting $x \rightarrow 0$, we obtain the boundary expressions

\begin{eqnarray}
d_x \overline{E}_I(s, 0) &=& -s \overline{E}_I(s, 0), \nonumber \\
d_x \overline{E}_R(s, 0) &=& s \overline{E}_R(s, 0).
\label{eq23.3}
\end{eqnarray}

The Laplace transform $\overline{E}_T(s, x)$ of the transmitted
field $E_T(t, x)$ satisfies the equation 

\begin{equation}
[d^2_{xx} 
- (s^2 + 1) ] \overline{E}_T(s, x) = 0, 
\label{eq24}
\end{equation}
which follows from (\ref{eq1.5}). We choose the causal solution (note
that $x>0$)

\begin{equation}
\overline{E}_T(s, x) 
= \overline{E}_T(s, 0) \exp[-(s^2 + 1)^{1/2} x], 
\label{eq25}
\end{equation}
so that, at the boundary $x=0$, we have
\begin{equation}
d_x \overline{E}_T(s, 0) 
= -(s^2 + 1)^{1/2} \overline{E}_T(s, 0).  
\label{eq26}
\end{equation}

Substitution of (\ref{eq23.3}) and (\ref{eq26}) into (\ref{eq21})
yields the boundary condition

\begin{equation}
s \overline{E}_I(s, 0)
- s \overline{E}_R(s, 0) 
= (s^2 + 1)^{1/2} \overline{E}_T(s, 0).
\label{eq27}
\end{equation}

Equations (\ref{eq21}) and (\ref{eq27}) imply that
\begin{eqnarray}
\overline{E}_R(s, 0) &=& 
\frac{s - (s^2 + 1)^{1/2}}{s + (s^2 + 1)^{1/2}} 
\overline{E}_I(s, 0), \nonumber \\
\overline{E}_T(s, 0) &=& 
\frac{2 s}{s + (s^2 + 1)^{1/2}} 
\overline{E}_I(s, 0).
\label{eq29}
\end{eqnarray}

It follows from the second of Eqs. (\ref{eq22.8}) that
 
\begin{equation}
\overline{E}_R(s, x) 
= \overline{E}_R(s, 0) \exp(s x).
\label{eq29.5}
\end{equation}

Finally, (\ref{eq25}), (\ref{eq29}), and (\ref{eq29.5}) yield

\begin{eqnarray}
\overline{E}_R(s, x) &=& 
\frac{s - (s^2 + 1)^{1/2}}{s + (s^2 + 1)^{1/2}} 
\exp(s x) \overline{E}_I(s, 0), \nonumber \\
\overline{E}_T(s, x) &=& 
\frac{2 s}{s + (s^2 + 1)^{1/2}} 
\exp[-(s^2 + 1)^{1/2} x] \nonumber \\
&\times& \overline{E}_I(s, 0).
\label{eq31}
\end{eqnarray}

The coefficients of $\overline{E}_I(s, 0)$ in (\ref{eq31})
are just the Green functions $\overline{\Gamma}_R(s, x)$ and
$\overline{\Gamma}_T(s, x)$ in Laplace space for the reflected and
transmitted pulse, respectively. We write the reflection Green function
in the form

\begin{equation}
\overline{\Gamma}_R(s, x) = \overline{R}(s) \overline{G}_R(s, x),
\label{eq32}
\end{equation}
where
\begin{eqnarray}
\overline{R}(s) &=& 
\frac{s - (s^2 + 1)^{1/2}}{s + (s^2 + 1)^{1/2}},
\nonumber \\
\overline{G}_R(s, x) &=& \exp(s x).
\label{eq34}
\end{eqnarray}
From the above discussion, it is clear that $\overline{R}(s)$
represents the reflection of the incident pulse at the vacuum-plasma
surface, whereas the factor $\overline{G}_R(s, x)$ accounts for
the subsequent propagation of the reflected pulse in vacuum.

Similarly, we write the transmission Green function in the form

\begin{equation}
\overline{\Gamma}_T(s, x) = \overline{T}(s) \overline{G}_T(s, x),
\label{eq35}
\end{equation}
where
\begin{eqnarray}
\overline{T}(s) &=& 
\frac{2 s}{s + (s^2 + 1)^{1/2}}, \nonumber \\
\overline{G}_T(s, x) 
&=& \exp[-(s^2 + 1)^{1/2} x].
\label{eq35.5}
\end{eqnarray}
Here $\overline{T}(s)$ represents the transmission of the incident
pulse across the vacuum-plasma surface, whereas the factor
$\overline{G}_T(s, x)$ represents the subsequent propagation of the
transmitted pulse in the plasma.

We see from (\ref{eq34}) and (\ref{eq35.5}) that $\overline{R}(s)$ and
$\overline{T}(s)$ are related through the equation

\begin{equation}
\overline{T}(s) = 1 + \overline{R}(s),
\label{eq36}
\end{equation}
which just states the fact that the electric field is conserved.

The influence of pulse duration and carrier frequency on the pulses'
transmission and subsequent propagation in a plasma can be investigated
by considering boundary fields $E_I(t,0)$ of the form

\begin{equation}
E_I(t, 0) = \exp(-\omega_e^2 t^2) \cos(\omega_c t).
\label{eq36.8}
\end{equation}
The parameters $\omega_e$ and $\omega_c$ are measures of the temporal
envelope width and carrier frequency respectively, of the incident
pulse at the $x=0$ boundary. We give in Table \ref{tab2} a
classification of the incident pulses (\ref{eq36.8}) at the boundary.

\begin{table}
\caption{Definition of the pulse classification scheme
employed in the text.}
\begin{tabular}{lc}
Pulse characteristic & Parameter regime \\
\tableline
Long duration (LD) & $\omega_e \ll 1$ \\
Intermediate duration (ID) & $\omega_e \simeq 1$ \\
Short duration (SD) & $\omega_e \gg 1$ \\
Low frequency (LF) & $\omega_c \ll 1$ \\
Intermediate frequency (IF) & $\omega_c \simeq 1$ \\
High frequency (HF) & $\omega_c \gg 1$ \\
\end{tabular}
\label{tab2}
\end{table}

The inverse temporal Laplace transform of (\ref{eq36}) is \cite{abr}
\begin{equation}
T(t) = \delta(t) - (2/t) J_2(t) H(t).
\label{eq37.5}
\end{equation}
$T(t)$ represents the part of the laser-plasma interaction in which
the incident pulse is transmitted across the vacuum-plasma boundary
$x=0$. The first term in (\ref{eq37.5}) represents the undistorted
transmission of a pulse into the plasma, while the second term
represents the reflection $R(t)$ at $x=0$, 

\begin{equation}
R(t) = - (2/t) J_2( t) H(t).
\label{eq37.6}
\end{equation}

This is evident by comparing (\ref{eq36}) with (\ref{eq37.5}).
Equation (\ref{eq37.6}) shows that the reflection of the laser pulse at
the vacuum-plasma boundary is not instantaneous, but rather a
decaying, oscillatory function of time. This indicates that there is a
harmonic response in the plasma to the incident pulse, which produces a
delayed, rather than instantaneous, reflected pulse. This response
takes the form of harmonic oscillations of plasma charges about their
equilibrium positions, which are induced by the incident sinusoidal
pulse.

\begin{figure}
\centerline{
\epsfxsize=0.6 \textwidth
\epsfbox[50 580 350 710]{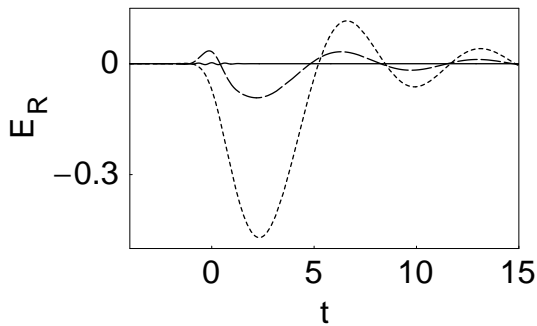} }
\caption{Temporal evolution of the reflection response $E_R(t, 0)$ [Eq.  (\protect\ref{eq37.8})] at the vacuum-plasma boundary, for
an incident pulse of intermediate duration ($\omega_e=1$) and carrier
frequency $\omega_c=1$ (dotted), $\omega_c=3$ (dashed), and
$\omega_c=10$ (solid).}
\label{fig1}
\end{figure}

\begin{figure}
\centerline{
\epsfxsize=0.6 \textwidth
\epsfbox[50 580 350 710]{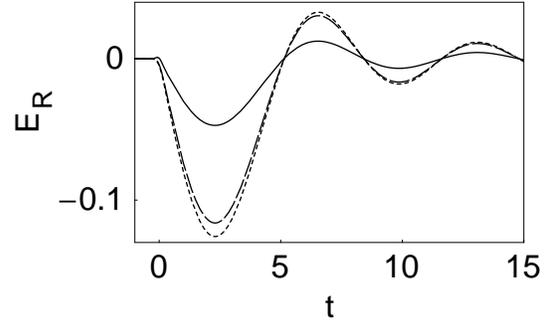} }
\caption{Same as in Fig.\protect\ref{fig1}, but for an incident pulse
of short duration ($\omega_e=5$).}
\label{fig2}
\end{figure}

One can investigate the dependence of the reflected pulse at $x = 0$ on
the duration and frequency of an impinging pulse $E_I(t, 0)$ by
calculating the convolution

\begin{equation}
E_R(t, 0) = \int_{-\infty}^\infty E_I(t', 0) R(t - t') \, dt'
\label{eq37.8}
\end{equation}
for different values of the parameters $\omega_e$ and $\omega_c$ in
$E_I(t,0)$, where $E_I(t,0)$ is given by (\ref{eq36.8}).  Figure
\ref{fig1} shows the reflection response for incident pulses of
intermediate duration (ID), with carrier frequencies varying from
intermediate (IF) to high (HF).  Figure \ref{fig2} shows the reflection
response for incident pulses of short duration (SD), again with carrier
frequencies varying from intermediate (IF) to high (HF).  It is seen in
Figs. \ref{fig1} and \ref{fig2} that the reflection response diminishes
as the carrier frequency of the pulse is increased. We also note that
as the duration of a pulse is shortened (i.e., as $\omega_e$ is
increased beyond $1$), the reflection response diminishes. This is
consistent with the fact that, as an incident pulse is shortened, more
of it will already have entered and propagated into the plasma before
the plasma's delayed reflection response [as described below
(\ref{eq37.6})] takes place. In particular, if $\omega_e \gg 1$,
the pulse is transmitted completely, with no distortion.

The propagation of the reflected pulse in vacuum is characterized by
the function $G_R(t, x)$, which is the inverse of $\overline{G}_R(s,
x)$ in (\ref{eq34}),

\begin{equation}
G_R(t, x) = \delta(t + x).
\label{eq37.7}
\end{equation} 
This means that the reflected pulse $E_R(t, x)$ has the space-time
dependence $E_R(t, x)=E_R(t + x)$, and propagates through the vacuum in
the negative $x$-direction away from the vacuum-plasma boundary, and
without distortion.

We now focus on the transmitted pulse $E_T(t, x)$. The propagation of
the transmitted pulse through the plasma is characterized by the
function $G_T(t, x)$ given by the inverse of $\overline{G}_T(s, x)$ in
(\ref{eq35.5}). $G_T(t, x)$ is found by first writing
$\overline{G}_T(s, x)$ as the spatial derivative

\begin{equation}
\overline{G}_T(s, x) = -\partial_x \overline{F}_T(s, x),
\label{eq38}
\end{equation}
where 
\begin{equation}
\overline{F}_T(s, x) 
= \frac{\exp[-(s^2 + 1)^{1/2} x]}{(s^2 + 1)^{1/2}}.
\label{eq38.3}
\end{equation}
The inverse of $\overline{F}_T(s, x)$ is given by \cite{abr}
\begin{equation}
F_T(t, x) = J_0[(t^2 - x^2)^{1/2}] H(t - x),
\label{eq38.5}
\end{equation} 
so that  
\begin{eqnarray}
G_T(t,x) &=& 
\delta(t-x) \nonumber \\
&-& x \frac{J_1[(t^2 - x^2)^{1/2}]}
{(t^2 - x^2)^{1/2}} H(t-x).
\label{eq38.8}
\end{eqnarray}
Equation (\ref{eq38.8}) represents the combined effect of a distortionless
propagation of the transmitted pulse (first term) and the propagation
of a dispersive wake generated by the plasma (second term).

We next compute the total Green function $\Gamma_T(t, x)$  by inverting
(\ref{eq35}). One way to do this is to compute $\Gamma_T(t, x)$ as the
convolution

\begin{equation}
\Gamma_T(t, x) =
\int_{-\infty}^\infty T(t-t') G_T(t', x) \, dt',
\label{eq42}
\end{equation}
where $T(t)$ is given by (\ref{eq37.5}), and $G_T(t,x)$ by
(\ref{eq38.8}). Again, (\ref{eq42}) clearly shows the two-stage process
of transmission followed by propagation. Analytic evaluation of
(\ref{eq42}) is quite involved. However, there is a simpler method for
obtaining $\Gamma_T(t, x)$ analytically that avoids integration, and
requires only the computation of derivatives. From (\ref{eq35}) and
(\ref{eq35.5}), we see that $\overline{\Gamma}_T(s, x)$ can be written
as the derivative

\begin{equation}
\overline{\Gamma}_T(s, x) = -2 \partial_x [s  f(s, x)], 
\label{eq44}
\end{equation}
where
\begin{equation}
f(s, x) = 
\frac{\exp[-(s^2 + 1)^{1/2} x]}
{[s + (s^2 + 1)^{1/2}] (s^2 + 1)^{1/2}}. 
\label{eq45}
\end{equation}
$f(s, x)$ has the inverse transform \cite{abr}
 
\begin{eqnarray}
F(t, x) &=& {\cal F}(t, x) H(t - x) =
\Biggl(\frac{t - x}{t + x}\Biggr)^{1/2} \nonumber \\
&\times& J_1[(t^2 - x^2)^{1/2}] H(t - x). 
\label{eq46}
\end{eqnarray}
This inverse transform only holds for $x > 0$, which is in accord with
our assumptions of the pulse entering the plasma at $x=0$, and
propagating into the plasma for $x > 0$. Since $F(0^+, x) = 0$ for $x
> 0$, we have from standard Laplace transform theory that $\partial_t
F(t, x)$ is the inverse transform of $s f(s, x)$. Therefore, from
(\ref{eq44}), we have

\begin{eqnarray} 
\Gamma_T(t, x) = -2 \partial_{tx}^2 F(t,x).  
\label{eq46.3} 
\end{eqnarray} 
The term $-2 \partial_{tx}^2 {\cal F}(t,x)$ in (\ref{eq46.3})
represents a modification to the incident pulse, caused by reflection
at the vacuum-plasma boundary and dispersion in the plasma. It is
given by

\begin{eqnarray}
-2 \partial_{tx}^2 {\cal F}(t,x) &=& 
- \frac{x t}{t + x} 
J_0(t^2 - x^2) \nonumber \\
&+& \Biggl(\frac{x t}{t + x} + 
\frac{2 (t -x)}{(t + x)^2}
\Biggr) \nonumber \\
&\times& J_2(t^2 - x^2).
\label{eq47}
\end{eqnarray}

From (\ref{eq31}), (\ref{eq35}), and (\ref{eq35.5}), we
see that the transmitted pulse $E_T(t, x)$ is given by the Green
function integral
\begin{equation}
E_T(t, x) = 
\int_{-\infty}^\infty E_I(t',0) \Gamma_T(t-t',x) \, dt'.
\label{eq51}
\end{equation}

\begin{figure}
\centerline{
\epsfxsize=0.6 \textwidth
\epsfbox[50 580 350 710]{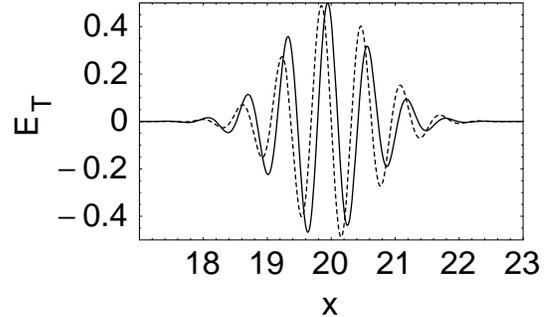} }
\caption{Spatial dependence of a transmitted pulse $E_T(20, x)$ [Eq.
(\protect\ref{eq51})] at time $t=20$ (solid). The incident pulse
crossed the vacuum-plasma interface at $t=0$, and had a spatial
dependence in vacuum characterized by Eq. (\protect\ref{eq36.8}),
$\omega_e = 1$ (ID), and $\omega_c = 10$ (HF). The incident pulse's
spatial dependence translated to $t=20$ is shown by the dotted curve,
for comparison with the resulting transmitted pulse.}
\label{fig3}
\end{figure}

\begin{figure}
\centerline{
\epsfxsize=0.6 \textwidth
\epsfbox[50 580 350 710]{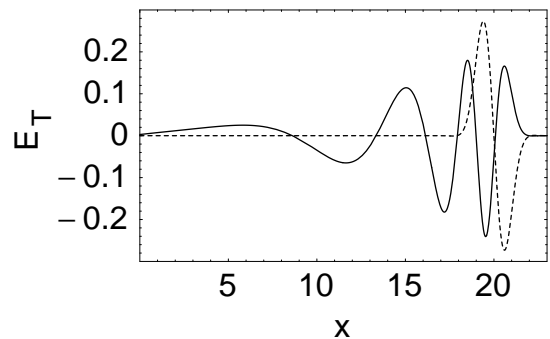} }
\caption{Same as in Fig.\protect\ref{fig3}, but with incident pulse
parameters $\omega_e = 1$ (ID) and $\omega_c = 1.5$ (IF).}
\label{fig4}
\end{figure}

We next perform the integration in (\ref{eq51}) for incident pulses of
the form (\ref{eq36.8}). We first consider incident pulses of
intermediate duration. Figures \ref{fig3} and \ref{fig4} show
plots of the propagation of ID-HF and ID-IF incident pulses,
respectively.

We see that the high frequency (HF) incident pulse propagates
practically undisturbed across the vacuum-plasma interface and into
the plasma, while the intermediate-frequency pulse develops an
electromagnetic (EM) wake. In the Appendix, we derive the following
perturbative expansions for $v_g$ and $v_p$ in the high-frequency case:

\begin{eqnarray}
v_g &\approx& 1 - \epsilon^2/2 - \epsilon^4/8, \nonumber \\
v_p = 1/v_g &\approx& 1 + \epsilon^2 /2 + 3 \epsilon^4 /8,
\label{eq52}
\end{eqnarray}
where $\epsilon = \omega_p / \omega_c$. The right sides of Eqs. 
(\ref{eq52}) are just the first three terms in the MacLaurin
expansions of $(1 - \epsilon^2)^{1/2}$ and $(1 - \epsilon^2)^{-1/2}$.

\begin{figure}
\centerline{
\epsfxsize=0.6 \textwidth
\epsfbox[50 580 350 710]{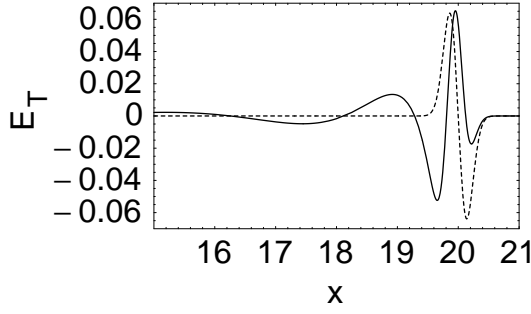} }
\caption{Same as in Fig.\protect\ref{fig3}, but with incident pulse
parameters $\omega_e = 5$ (SD) and $\omega_c = 1.5$ (IF).}
\label{fig5}
\end{figure}

We next consider the propagation of short (SD) incident pulses. Figure
\ref{fig5} shows a plot of an incident SD-IF pulse.

As expected, the wake generation is smaller than for the incident
ID-IF pulse. And as the frequency of the incident SD pulse is
increased, it is found that wake generation is practically
nonexistent.

\section{Summary}

In this paper, we considered the transmission and reflection of an
electromagnetic pulse at a vacuum-plasma boundary, and the subsequent
propagation of the transmitted pulse in the plasma. We extended the
well-known theory for long pulses into the short-pulse regime, in
which the pulse duration is comparable to the inverse plasma frequency.
When the carrier frequency of the incident pulse is much higher than
the plasma frequency, most of the \mbox{incident} pulse is transmitted
without distortion. Subsequently, the transmitted pulse propagates
without distortion at \mbox{its} group speed. When the carrier
frequency is comparable to the plasma frequency, the transmitted pulse
is distorted, and leaves behind it an electromagnetic wake. The
reflected pulse is delayed relative to the incident pulse, and is also
distorted. When the carrier frequency is less than the plasma
frequency, the incident pulse is absorbed by the plasma before being
reemitted.

\acknowledgements

This work was supported by the National Science Foundation under
Contract No. PHY94-15583, the Department of Energy (DOE) Office of
Inertial \mbox{Confinement} Fusion under Cooperative Agreement No.
DE-FC03-92SF19460, the University of Rochester, and the New York State
Energy Research and \mbox{Development} Authority. 

\appendix
\section{Propagation of a high-frequency pulse}

Let $\omega_c t \rightarrow t$, $\omega_c x/ c \rightarrow x$, and
$\omega_p / \omega_c \rightarrow \epsilon$, so that $t$ and $x$ 
become dimensionless. Then the wave equation (\ref{eq1}) can be written
as

\begin{equation}
\bigl(\partial^2_{tt} - \partial^2_{xx} + \epsilon^2\bigr)E = 0.
\label{eqa1}
\end{equation}
The study of pulse propagation is facilitated by the characteristic
transformation

\begin{equation}
\tau = t - \beta x, \ \ \xi = x - \beta t, 
\label{eqa2}
\end{equation}
where $\beta < 1$. In terms of the characteristic variables $\tau$ and
$\xi$, the wave equation (\ref{eqa1}) can be rewritten as

\begin{equation}
\bigl[(1 - \beta^2)(\partial^2_{\tau\tau} - \partial^2_{\xi\xi})
+ \epsilon^2\bigr]E = 0. 
\label{eqa3}
\end{equation}

One can solve Eq. (\ref{eqa3}) by using multiple scale analysis
\cite{nay}. To do this, one introduces the time and distance scales

\begin{equation}
\tau_n = \epsilon^n\tau, \ \ 
\xi_n = \epsilon^n\xi.
\label{eqa4}
\end{equation}

Correct to second order, one can write

\begin{eqnarray}
\partial_\tau &\approx& \partial_{\tau_0} + \epsilon\partial_{\tau_1}
+ \epsilon^2\partial_{\tau_2}, \nonumber \\ 
\partial_\xi &\approx& \partial_{\xi_0} + \epsilon\partial_{\xi_1}
+ \epsilon^2\partial_{\xi_2}. 
\label{eqa5}
\end{eqnarray}
Guided by the well-known characteristics of a long pulse, we assume
that

\begin{equation}
\beta \approx 1 + \epsilon^2\beta_2 + \epsilon^4\beta_4 
\label{eqa6}
\end{equation}
and

\begin{equation}
E(\tau,\xi) = B(\tau_2,\xi_1)\exp(-i\tau_0). 
\label{eqa7}
\end{equation}
Ansatz (\ref{eqa7}) corresponds to a pulse that has a carrier frequency
of unity and an amplitude that varies on the \mbox{slow} scale $\xi_1$. For
this amplitude variation, $\beta$ is the group speed of the pulse, and
the characteristic variables are proportional to time and distance
measured in the pulse frame. One now substitutes Eqs. (\ref{eqa5}) -
(\ref{eqa7}) in Eq. (\ref{eqa3}) and collects terms of like order.
The zeroth- and first-order equations are satisfied automatically
by construction.

In second order,

\begin{equation}
\bigl[-2\beta_2(\partial^2_{\tau_0\tau_0} - \partial^2_{\xi_0\xi_0})
+ 1\bigr]E = 0. 
\label{eqa8}
\end{equation}
It follows from Eq. (\ref{eqa8}) and ansatz (\ref{eqa7}) that

\begin{equation}
\beta_2 = -1/2. 
\label{eqa9}
\end{equation}
In third order,
\begin{equation}
-4\beta_2 (\partial^2_{\tau_0\tau_1}
 - \partial^2_{\xi_0\xi_1})E = 0. 
\label{eqa10}
\end{equation}
Equation (\ref{eqa10}) is consistent with ansatz (\ref{eqa7}), in which
$E$ is assumed to be independent of $\xi_0$ and $\tau_1$. In fourth
order,

\begin{eqnarray}
&-&\bigl[4\beta_2(\partial^2_{\tau_0\tau_2} - \partial^2_{\xi_0\xi_2})
+2\beta_2(\partial^2_{\tau_1\tau_1} - \partial^2_{\xi_1\xi_1}) 
\nonumber\\ 
&+&
(2\beta_4 + \beta_2^2)(\partial^2_{\tau_0\tau_0} -
\partial^2_{\xi_0\xi_0}) \bigr]E = 0.
\label{eqa11}
\end{eqnarray}
The pulse has a carrier frequency of unity by construction, so the
dependence of $E$ on $\tau_2$ cannot be oscillatory. It follows from
this constraint that $(2\beta_4 + \beta_2^2) = 0$ and, hence, that

\begin{equation}
\beta_4 = -1/8. 
\label{eqa12}
\end{equation}
The group speed $\beta \approx 1 - \epsilon^2/2 - \epsilon^4/8$, which
is just the first three terms in the Maclaurin expansion of $(1 -
\epsilon^2)^{1/2}$. The remaining nonzero terms in Eq. (\ref{eqa11}) are

\begin{equation}
\bigl(2i\partial_{\tau_2} + \partial^2_{\xi_1\xi_1}\bigr)B = 0,
\label{eqa13}
\end{equation}
which describe the dispersal of the pulse.

Finally, note that ansatz (\ref{eqa7}) constrains the phase speed to be
the inverse of the group speed. Since no contradictions appear in the
subsequent analysis, the assumptions underlying ansatz (\ref{eqa7}) are
correct. One can also use the ansatz

\begin{equation}
E(\tau,\xi) = B(\tau_2,\xi_1)
\exp[i\nu\xi_0 - i(1 - \nu\beta)\tau_0], 
\label{eqa14}
\end{equation}
which does not constrain the phase speed, but leads to the same result.

\end{document}